\documentclass[reprint,amsmath,amssymb,aps,linenumbers]{revtex4-1}
\usepackage[utf8]{inputenc}
\usepackage{CJK}

\usepackage{url}
\usepackage{hyperref}

\usepackage{graphicx}
\usepackage{dcolumn} 
\usepackage{bm} 
\usepackage{physics}

\usepackage{comment}

\usepackage{wasysym} 
\begin{document}
\nolinenumbers
\begin{CJK*}{GB}{}

\title{How Social Network Structure Impacts the Ability of Zealots to Promote Weak Opinions}

\author{Thomas Tunstall}
\altaffiliation{Living Systems Institute, Faculty of Health and Life Sciences, University of Exeter}
\altaffiliation{Physics and Astronomy, Faculty of Environment, Science and Economy, University of Exeter}
\altaffiliation{Mathematics and Statistics, Faculty of Environment, Science and Economy, University of Exeter}

\date{\today}

\begin{abstract}
Social networks are often permeated by agents who promote their opinions without allowing for their own mind to be changed: Understanding how these so-called `zealots' act to increase the prevalence of their promoted opinion over the network is important for understanding opinion dynamics. In this work, we consider these promoted opinions to be `weak' and therefore less likely to be accepted relative to the default opinion in the network. We show how the proportion of zealots in the network, the relative strength of the weak opinion, and the structure of the network impact the long-term proportion of the those in the network who subscribe to the weak opinion.
\end{abstract}

\maketitle
\end{CJK*}

\section{Introduction}


Whether it be deciding what to buy, who to vote for, or what to believe, one's opinions are shaped by the opinions of those they interact with. In order to model the manner in which opinions change over time in a social network, the voter model \cite{clifford1973model,holley1975ergodic} is one of the simplest models employed. In this model, each node in the network corresponds to an individual who can adopt one of a variety of opinions. During each simulation step, a random individual is selected to adopt the opinion of a random neighbour; this process is usually repeated until `consensus' is achieved when each individual has adopted the same opinion. The time taken until consensus is achieved has been the focus of studies into the voter model, as has the transient properties of interfaces (`active links') between nodes subscribing to different opinions for a variety of different types of graphs (from $N$-dimensional lattice graphs \cite{krapivsky1992kinetics,frachebourg1996exact} to random networks \cite{castellano2003incomplete,sood2005voter,castellano2009statistical,yildiz2010voting}). There are many modifications to the original voter model in order to more accurately represent the manner in which individual's opinions are shaped by their interactions with others: A review in this context is presented in Ref.\ \cite{redner2019reality}. 

In real social networks, not all opinions may be accepted equally: Some may correspond to facts or concepts which the layperson may find counter-intuitive (for example, abstract scientific ideas), some may conflict with deeply-held biases of the individual (for example, religious or political affiliation), or some may be suppressed by societal pressure (for example, the expectation to conform to gender or cultural norms). In such cases, a more accurate way to model opinion dynamics may be to modify the voter model to incorporate a `weaker' opinion which is less likely to be adopted than a default stronger opinion when both are considered. The literature which emphasises this focuses on the time taken for consensus to be achieved \cite{tang2015evolutionary,mukhopadhyay2020voter}, and the probability that a consensus to the stronger opinion will occur given the original placement of a single stronger opinion on a random graph \cite{antal2006evolutionary,sood2008voter}. 


Not all individuals may change their mind in the same way: For example, it could be that an individual holds to a political or religious belief or idea with such conviction that their mind cannot easily be changed. In the voter model literature such individuals are defined as `zealots' \cite{mobilia2003does}, of which there are two varieties: `flexible' zealots \cite{mobilia2003does,mobilia2005voting,caligiuri2023noisy} whose minds can be changed yet prefer a particular opinion, and `inflexible' zealots \cite{mobilia2007role,mobilia2015nonlinear} who never change their minds from their preferred opinion. The impact of a single zealot \cite{mobilia2003does} and multiple zealots of each opinion \cite{mobilia2005voting,mobilia2007role,mobilia2015nonlinear,caligiuri2023noisy} on the long-term magnetization (the proportion of edges which are shared by nodes of differing opinions) on the graph has been investigated, even outside the context of voters on a network \cite{galam2007role}. Studying how zealots impact opinion diversity in a group is an active area of study, with respect to understanding opinion control \cite{vendeville2022towards,moeinifar2021zealots,czaplicka2022biased}, the spread of political opinions \cite{braha2017voting}, and vaccine hesitancy \cite{muller2022echo}. Here we are interested in scenarios in which individuals or organizations specifically aim to disseminate specific ideas without engaging at all with the other opinion: Advertisements, propagandists and conspiracy theorists are suitable examples. Therefore, in this work we shall consider only the presence of inflexible zealots.

Despite the wealth of literature addressing the implementation of weaker opinions and zealots separately in a voter model, there has been no investigation into the implementation of both at the same time. In a world saturated with commercial and political advertising, it is of interest to determine how well a finite proportion of zealots which promote a weak opinion can influence opinion diversity over the network. 
Human social networks are typically long-ranged and exhibit a high degree of clustering  \cite{barrat2008dynamical,castellano2009statistical}, with most members of a community being a member of a giant component of the social network \cite{newman2001structure}. As a consequence of this, modelling practices often take these networks to be complete graphs, referred to as the `mean-field' approach \cite{braha2017voting}. Separately, it is also appreciated that the degree distribution on random graphs can have significant effects on the dynamics on the graph \cite{castellano2009statistical}. To account for both of these regimes, we shall consider the network to be of the Erd{\H{o}}s - R{\'e}nyi variety, \cite{erdds1959random} where we can tune the mean degree of the graph, $C$, to transition from a complete graph (for $C=N-1$, where $N$ is the number of nodes on the graph) to a collection of isolated trees (for $C<1$). This choice is also made due of the wealth of literature regarding the evolution of connected components in an Erd{\H{o}}s - R{\'e}nyi graph as $C$ is varied - this work is otherwise agnostic with respect to the choice of random graph variety. Here we study the long-time diversity of opinions in the presence of zealots which promote the weaker opinion only across an Erd{\H{o}}s - R{\'e}nyi graph, in order to provide an insight into how the ability of zealots to propagate a weaker opinion is impacted by the structure of the community.

Although the model presented in this work is primarily framed in the context of opinion diversity in a social network, it can also be used to describe evolution in the game-theoretic or ecological contexts. In the evolutionary game theory context, the voter model is analogous to the Moran model \cite{moran1958random} on a directed graph \cite{nowak2006evolutionary}. In this sense, opinion strength could corresponds to the effect of selection, and zealots correspond to `root' nodes in which the all edges are directed away from them \cite{nowak2006evolutionary}. This is also closely related to game-theoretic approaches which implement a payoff matrix which incentivizes a `defector' strategy, and zealots correspond to individuals who hold rigidly to the weaker `cooperator' strategy regardless: It has been illustrated that in this case the zealots can dramatically promote the cooperation strategy in both well-mixed \cite{masuda2012evolution,cardillo2020critical} and network-based scenarios \cite{cardillo2020critical}.

In the ecological evolution context, nodes may instead correspond to environments over which sub-populations of organisms can survive, with edges corresponding to migratory pathways. Here, zealots correspond to environments in which there is a selection pressure which promotes only a sub-population which otherwise exhibits a `cost to resistance' \cite{andersson_biological_1999,kliot2012fitness} outside of that environment (thus corresponding to the weak opinion). In this context, the evolution of the modified voter model simulates a coarse-grained description of how heterogeneously distributed selection pressure promotes genetic diversity. Therefore, this model would be useful for determining how spatial heterogeneity in control measure application (such as pesticides) can be distributed to prevent a resistant sub-population from rising to fixation.

\begin{figure}
    \centering
    \includegraphics{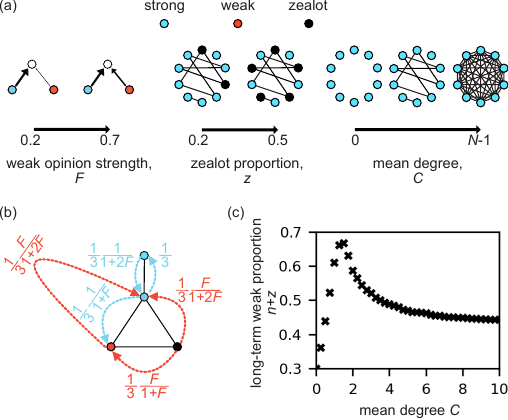}
    \caption{ (a) Illustration of the effects of changing relative weak opinion strength, $F$, the proportion of the network who are zealots, $z$, and the mean degree of nodes in an Erd{\H{o}}s - R{\'e}nyi graph, $C$. (b) Underlying modified voter model applied to an example graph. The possible updates during the next simulation step are illustrated by dashed arrows, annotated with the corresponding probability of each event occurring. (c) The mean long-term frequency of the weak opinion over the entire graph (of size $N=10^4$) for varying $C$, for $F=0.3$ and $z=0.3$, averaged over the last $10\%$ of $10^9$ simulations steps of $10$ repeats of the modified voter model. All free nodes are initially of the strong variety.}
    \label{Fig1}
\end{figure}

\section{Model \& Methods}
Consider a graph of $N$ nodes, where each node adopts either opinion $W$ (`weaker') or $S$ (`stronger'). A number of nodes, $Z$, are zealots which subscribe to the weaker opinion only. Therefore the number of nodes `free' to change their opinion is $N_{\text{free}}= N-Z$. The number of free nodes subscribed to $W$ and $S$ at any given time are denoted by $N_W$ and $N_S$ respectively: As $N_W + N_S = N_{\text{free}}$, the state of the system will be described entirely by $N_W$. 

At each simulation step, a random node is chosen equiprobably from the free nodes to adopt the opinion of one of its neighbours at random. The probability that a $W$ neighbour is selected during this step is reduced by a `fitness' factor $F: 0<F\leq1$ relative to a $S$ neighbour. Therefore, the probability of adopting the $W$ opinion given $a$ neighbours of the $W$ opinion and $b$ neighbours of the $S$ opinion is $F a / (F a+b)$. An example state of a small network is demonstrated in Fig.\ \ref{Fig1}b, in which the arrows correspond to all the possible update events in the next simulation step, notated with the probability of that transition occurring.

We choose the underlying random graph to be of the Erd{\H{o}}s - R{\'e}nyi variety, in which the probability that an edge exists between two nodes is $p = C/(N-1)$, independent of edges between other nodes. When $C = N-1$ the Erd{\H{o}}s - R{\'e}nyi graph corresponds to the complete graph, and thus in this regime we can obtain the corresponding `mean-field' result by employing a master equation using a similar approach to as in Ref.\ \cite{mobilia2007role}. For $C<1$ the Erd{\H{o}}s - R{\'e}nyi graph consists of many tree-like connected components \cite{erdHos1960evolution,bollobas_random_2001}.

Of particular interest to us is the long-term distribution of opinions across the graph: We denote the fraction of nodes which are zealots as $z=Z/N$ and the proportion of the nodes which are free and subscribe to the weak opinion as $n=N_W/N$. In Fig.\ \ref{Fig1}c we show the results of a simulation of the modified voter model on a network of size $N=10^4$ for $10^9$ simulation steps, and determine the long-term mean of the overall proportion of the weak opinion, $n+z$. To form an analytical description of these example results, we shall consider the complete-graph limit of the Erd{\H{o}}s - R{\'e}nyi graph ($C\rightarrow N-1$), and then use our understanding of trees to create an analytical prediction for the $C<1/(1-z)$ regime. We shall then discern how to interpolate between these extremes to provide a heuristically useful approximation. 

\section{The Mean-Field Limit}

The steady-state proportion of nodes which subscribe to the weaker opinion across a finite complete graph of size $N$, $P(n;N,F,z)$, can be obtained by solving the Fokker-Plank equation in a similar manner presented in Ref.\ \cite{mobilia2007role}, except for the implementation of a relative fitness factor $F$ which reduces the probability that a weaker opinion neighbour is selected. We set a fixed proportion, $z$, of the nodes to be zealots of the weaker opinion. Furthermore, we set a single node to be a zealot of the stronger opinion: $z_S=1/N$: This is to help simulate an infinite network in the mean-field limit, ensuring that the absorbing state cannot be entered in finite time. The solution to the corresponding Fokker-Plank equation in the limit that $N\to\infty$ (such that $z_S\to0$) is therefore (see Supplementary Material \cite{Supplement}): 

\begin{equation}\label{SteadyState}
    \begin{split}
        &P(n;N,F,z) = \mathcal{Z} \frac{\exp{2 \int_0^n dn' \alpha(n')/\beta(n')}}{\beta(n)},\\
        &\alpha(n;N,F,z)= \frac{1-z-n}{1-z}\frac{F(n+z) -n}{1-z-n+F(n+z)},\\
        &\beta(n;N,F,z)=  \frac{1-z-n}{(1-z)N}\frac{F(n+z) + n}{1-z-n+F(n+z)},
    \end{split}
\end{equation}

in which $\mathcal{Z}$ is a normalization factor. 
From here, we can deduce the deterministic steady state proportion of the free resistant opinion, $n^*(F,z)$, and define the steady state total proportion of nodes subscribing to the weak opinion $t_{MF}(F,z) = n^*(F,z) + z$ (see Supplementary Material \cite{Supplement}):

\begin{equation}\label{CompleteGraph}
   t_{MF}(F,z) = \min{\left(\frac{z}{1-F},1\right)}.
\end{equation}

We shall refer to the case where $t_{MF}(F,z) = 1$ as the network being supercritical.

\begin{figure}[h!]
    \centering
    \includegraphics{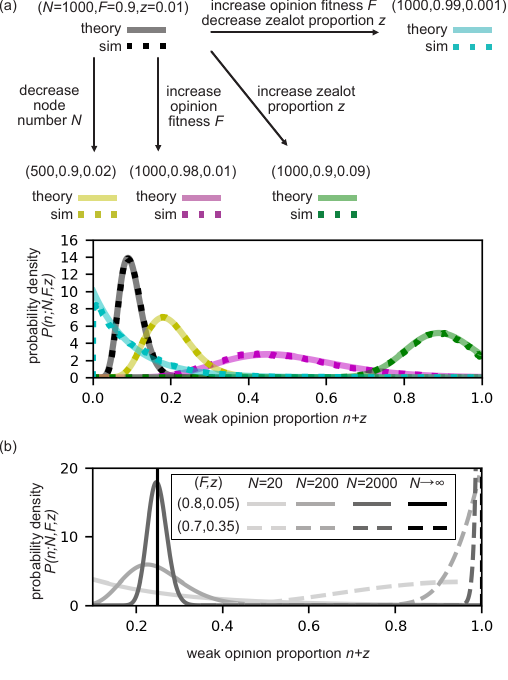}
    \caption{(a) Comparison between the analytical steady-state solution given by Eqn.\ \ref{SteadyState} (solid line) and the normalised frequency density of a simulation (dotted lines) for a variety of $(N,F,Z)$ combinations over $10^8$ simulation steps. Starting with $(1000,0.9,10)$ (black) we see how increasing $F$ (magenta), increasing $Z$ (green) or decreasing $N$ (yellow) acts to increase the overall long-term proportion of weak opinion. Decreasing $Z$ and increasing $F$ to preserve the equilibrium number of nodes subscribed to the weak opinion (given by Eqn.\ \ref{CompleteGraph}) is also compared (cyan). (b) Analytical steady-state solution given by Eqn.\ref{SteadyState} for a varying number of nodes, $N$, for the cases of $(F,z) = (0.8,0.05)$ (solid) and $(F,z) = (0.7,0.35)$ (dashed). The $N\rightarrow\infty$ vertical lines corresponds to the results of Eqn.\ \ref{CompleteGraph}. }
    \label{Fig2}
\end{figure}

The impact of varying $N$, $F$, and $z$ on the distribution of the long-term proportion of nodes subscribing to the weak opinion on a finite network is visualised in Fig.\ \ref{Fig2}a: Starting with a base case of $(N=1000,F=0.9,z=0.01)$ (black), we compare the long-term proportion of the weak opinion over the population as predicted in Eqn.\ \ref{SteadyState} (solid lines) against experiment (dotted lines), seeing strong agreement. The ability of zealots to increase the proportion of the weak opinion can be improved by adding more zealots ($z: 0.01 \rightarrow 0.09$, green), increasing the weak opinion fitness ($F: 0.9\rightarrow0.98$, magenta) or by being present in a smaller network ($N: 1000\rightarrow500$, yellow). $(N=1000,F=0.99,z=0.001)$ (cyan) corresponds to decreasing $z$ and increasing $F$ in order to preserve the steady-state proportion given in Eqn.\ \ref{CompleteGraph}: This highlights the ability of even a single zealot to prompt a finite proportion of nodes to adopt the weak opinion in a finite well-mixed social network.

In Fig.\ref{Fig2}b we illustrate that the larger the number of nodes in the network (given by a darker gray line color), the closer the peak of the analytical distribution given by Eqn.\ \ref{SteadyState} gets to the prediction of the infinite mean-field limit given by Eqn.\ \ref{CompleteGraph} (the black vertical lines) for a subcritical (solid line) and supercritical (dashed line) example.


\section{Trees}

As $N\rightarrow\infty$, it is known that in an Erd{\H{o}}s - R{\'e}nyi graph with mean degree $C>1$ a giant component exists which contains a positive fraction, $g(C)$ of all nodes, where \cite{tishby2018revealing} 

\begin{equation}\label{g}
    g(C) = 1 + \frac{W(-C e^{-C})}{C},
\end{equation}

in which $W(x)$ is the Lambert W-function. The remaining nodes belong to trees, where the expected number of trees of size $\mathcal{S}$ in a graph of size $N$ is \cite{bollobas_random_2001}

\begin{equation}\label{TreeNumDist}
    \begin{split}
        \mathcal{N}(\mathcal{S};C,N) &= \binom{N}{\mathcal{S}} \mathcal{S}^{\mathcal{S}-2} \left(\frac{C}{N}\right)^{\mathcal{S}-1} \\
        & \times \left(1-\frac{C}{N}\right)^{\binom{\mathcal{S}}{2}-(\mathcal{S}-1)+\mathcal{S}(N-\mathcal{S})}.
    \end{split}
\end{equation}

The presence of a lone zealot within a finite tree has much more profound consequences than in the infinite complete graph: The zealot now acts as the source of a random walk of the weaker opinion, which is biased in the direction of the zealot. In finite time the weaker opinion will eventually reach all ends of the tree, resulting in every node within the tree adopting the weak opinion with no possibility of reverting to the strong opinion.

\begin{figure}
    \centering
    \includegraphics{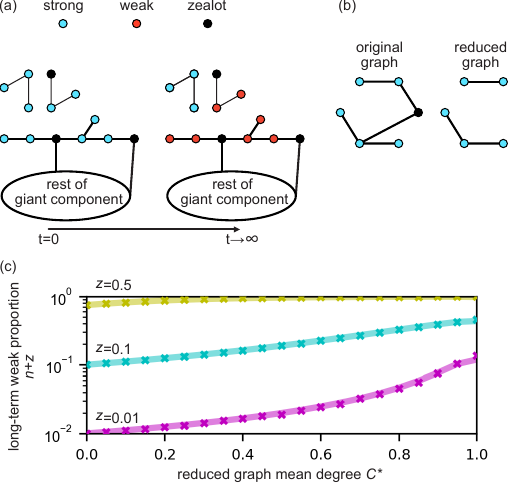}
    \caption{(a) The long-term fate of trees containing a zealot and tree-like subgraphs bound by zealots in the giant component is consensus to the weak opinion. (b) Visualization of how removing zealots from a graph creates a reduced graph. (c) How simulation (crosses) and theory (solid line, given by Eqn.\ \ref{ER_SUPERCRITICAL_TREES}) agree for the long-term proportion of weak opinion over a graph of size $N=10^4$ as a function of $C^*$ for $z = 0.001$ (magenta), $z = 0.1$ (cyan), $z = 0.5$ (yellow). In all cases, $F=1$, and the simulation results are the median of $100$ repetitions of the steady state obtained.}
    \label{Fig3}
\end{figure}

The presence of zealots in the giant component may result in the existence of subgraphs for which there are no paths to any free node outside the subgraph without passing a zealot node. These subgraphs are trees in the limit of $N\rightarrow\infty$, and therefore will go on to entirely adopt the weak opinion in finite time (see Fig.\ \ref{Fig3}a). One can account for both the trees outside and embedded within the giant component by removing all zealots from the network (see Fig.\ \ref{Fig3}b), and ascertaining the resultant distribution of trees using Eqn.\ \ref{TreeNumDist}. From here we can determine the expected frequency of trees which had shared at least one edge with a zealot, and assert that all nodes within such a tree becomes of the weak opinion in the long term. The behavior of the remaining giant component (if one exists) will be discussed in the next section.

We construct the graph of $N$ nodes of mean degree $C$ (such that the probability of any two nodes sharing an edge is $p=C/N$). Of these nodes, a proportion $z$ are zealots. After removing the zealots, we are left with a `reduced' graph of size $N^*=(1-z)N$, and mean degree $C^* = (1-z)C$: Note that the probability of two nodes sharing an edge, $p$, is unchanged, and the graph is still of the Erd{\H{o}}s - R{\'e}nyi variety. The number of trees of size $\mathcal{S}$ can be evaluated to be $\mathcal{N}(\mathcal{S};C^*,N^*)$  (Eqn.\ \ref{TreeNumDist}). For a tree of size $\mathcal{S}$ in the reduced graph, the probability than an individual node does not share an edge with a zealot in the original network is $(1-p)^{Nz}$.
The corresponding probability that none of the nodes within the reduced graph tree independently share an edge with a zealot in the original network is $(1-p)^{Nz\mathcal{S}}$. Therefore, the probability that the tree shares at least one edge with a zealot is $1-(1-p)^{Nz\mathcal{S}}$. The expected contribution of long-term weak opinion from trees of size $\mathcal{S}$ is therefore $\mathcal{N}(\mathcal{S};C^*,N^*)\times \mathcal{S} \times \left[1-(1-p)^{Nz\mathcal{S}}\right]$. The total long-term proportion of nodes over the entire original network (where the zealots are present) which subscribe to the weak opinion as a result of being in these kinds of trees is therefore:

\begin{equation}\label{ER_SUPERCRITICAL_TREES}
\begin{split}
    &t_{\text{tree}}(C^*,N^*,z) =\\
    &\frac{1}{N} \sum_{\mathcal{S}=1}^{N^*} \mathcal{N}(\mathcal{S};C^*,N^*)\times \mathcal{S} \times \left[1-(1-p)^{Nz\mathcal{S}}\right].
\end{split}
\end{equation}

Eqn.\ \ref{ER_SUPERCRITICAL_TREES} accounts for trees in the reduced graph, so we must ask if a giant component exists in the reduced graph. Using Eqn.\ \ref{g} we know that a giant component will only emerge if $C^* > 1$. Therefore, if $(1-z)C <1$ we expect Eqn.\ \ref{ER_SUPERCRITICAL_TREES} to fully describe the long-term proportion of weak opinion of the free nodes: This theory is compared against direct simulation for $z=0.01, 0.1, 0.5$ in Fig.\ \ref{Fig3}c to excellent agreement.

\section{Treatment of the Giant Component}

The results for $C<1/(1-z)$ are exact, but for $C\geq1/(1-z)$ a giant component emerges in the reduced graph (a `reduced' giant component) which is difficult to describe exactly. We can create a heuristically accurate approximation by assuming that the reduced giant component matches the behaviour of a complete graph.
To this end, we need to evaluate the expected proportion of zealots which will share at least one edge with the reduced giant component. For a given zealot, the probability that it does not share an edge with any of the $N g(C^*)$ non-zealot nodes in the reduced giant component is $(1-p)^{N g(C^*)}$. This sets up a binomial problem where the expectation number of zealots which share an edge with the reduced giant component is $N z \left[1-(1-p)^{N g(C^*)}\right]$. This newly constructed component has an effective zealot proportion, $z^*$:

\begin{equation*}
    z^* = \frac{z \left[1-(1-p)^{N g(C^*)}\right]}{(1-z)g(C^*) + z\left[1-(1-p)^{N g(C^*)}\right]}.
\end{equation*}

If we assume the giant component matches the mean-field prediction, the proportion of all nodes in the giant component which subscribe to the weak opinion in the $N\rightarrow\infty$ limit is given by Eqn.\ \ref{CompleteGraph}, using $z^*$ as the proportion of zealots. To obtain the number of free nodes subscribed to the weak opinion, we subtract the expected number of zealots in this giant component: We therefore predict the long-term proportion of nodes in the entire network which subscribe to the weak opinion as a result of being within the effective giant component is defined as $t_{\text{GC}}(C^*,N^*,z)$:

\begin{equation}\label{ER_GiantComponentContribution}
    \begin{split}
        &t_{\text{GC}}(C^*,N^*,z,F) = \\
        &\left((1-z)g(C^*) + z\left[1-(1-p)^{N g(C^*)}\right]\right)t_{MF}(F,z^*)\\
        & - z \left[1-(1-p)^{N g(C^*)}\right].\\
    \end{split} 
\end{equation}

 The total long-term proportion of weak nodes over the entire original network is therefore the contribution from trees in the reduced graph (Eqn.\ \ref{ER_SUPERCRITICAL_TREES}), the contribution from the giant component in the reduced graph (Eqn.\ \ref{ER_GiantComponentContribution}), and the contribution from the zealots themselves:

\begin{equation}\label{FULL}
    \begin{split}
        t_{\text{tot}}(C,N,z,F) =& t_{\text{tree}}(C^*,N^*,z) + t_{\text{GC}}(C^*,N^*,z,F) +z.
    \end{split}
\end{equation}

\begin{figure}
    \centering
    \includegraphics{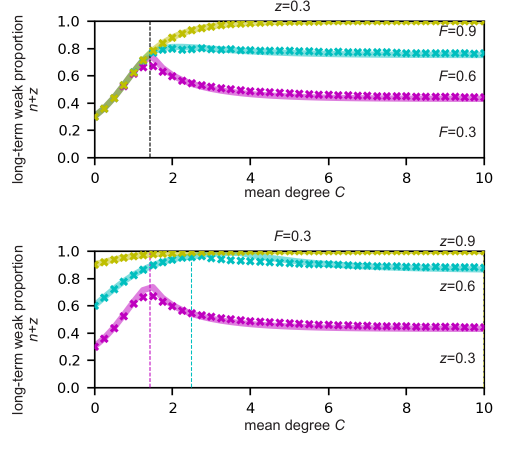}
    \caption{How simulation (crosses) and theory (solid line, given by Eqn.\ \ref{FULL}) agree for constant $z=0.3$ and different $F$ (top) and for constant $F=0.3$ and different $z$ (bottom), each for varying $C$. Vertical dashed lines correspond to $C=1/(1-z)$ for each value of $z$: These are identical for the upper graph and given in black. Simulation results are averaged over the last $10\%$ of $10^9$ simulations steps of $10$ repeats of the modified voter model on a network of size $N=10^4$. All free nodes are initially of the strong variety.}
    \label{Fig4}
\end{figure}

This result is compared against simulation in Fig.\ \ref{Fig4}, where we see the approximation works best for $C<1/(1-z)$, as this is where $g(C^*) = 0$ and our exact analytical argument of Eqn.\ \ref{ER_SUPERCRITICAL_TREES} is valid. For $C\geq1/(1-z)$ we see Eqn.\ \ref{FULL} is often close to the simulated results, suggesting that there are some cases where the zealots propagate the weak opinion differently in the non-complete regime than in the complete regime. Note for the $z=0.3$ case shown in Fig.\ \ref{Fig4} (upper), the theoretical predictions are identical for the tree-like limit ($C\leq 1/(1-z)$), diverging only when a giant component emerges (after the black dashed line) in the reduced graph and the fitness has an impact on opinion diversity.

\section{Conclusion \& Discussion}

We have demonstrated how the incorporation of a finite proportion of zealots subscribing to a weak opinion can shift opinion diversity over random networks. We analytically derived the long-term proportion of nodes subscribing to the weak opinion in both the mean-field limit by studying the corresponding Fokker-Planck equation, and for $C<1/(1-z)$, where the results are determined by the distribution of trees in a reduced graph obtained by removing all zealots from the original network. We also provided a heuristically useful approximation for how this proportion changes in the intermediate range $1/(1-z) < C < N-1$. Of particular interest is the observation that this proportion can be maximised by satisfying the conditions for supercriticality in the mean-field limit.

There are a variety of applications for this result, which inform different future avenues of research. It follows from Eqn.\ \ref{CompleteGraph} that for a given number of zealots in a network, there is a critical fitness, $F^* = 1-z$, of their promoted opinion above which all connected components are likely to become dominated by this opinion. In this scenario, the initially weak opinion of fitness $F\geq F^*$ has now become consensus. At this point, the zealots could switch to promote an opinion of fitness $F$ relative to the previous weak opinion (or fitness $F^2$ relative to the original opinion), and this new weaker opinion could then go on to become consensus. Repeating this process can result in a consensus opinion with relative fitness much less than $F^*$. This process by which a social network can be slowly converted due to variable zealots would be a powerful tool for controlling opinion dynamics to a degree that static zealots cannot, and would be an interesting avenue for future research.

In the context of advertising, we have provided a useful heuristic model which can be used to predict how a change in the proportion of zealots, $z$, for a given $F$ and $C$ can shift the proportion of individuals who subscribe to the weak opinion. This can provide the basis for a cost-benefit analysis to optimize advertising practice: Does the expected returns from converting a given number of people to the weak opinion outweigh the cost of implementing/hiring more zealots? 

Finally, in the case of top-down opinion control by changing connectivity $C$, we see in Fig.\ \ref{Fig4} there may be cases in which there is a global maximum in the proportion of weak opinion. This fact can be used to promote or suppress a weak opinion promoted by zealots by controlling how the population interacts. 

To better describe realistic social network dynamics, the underlying model may need to be adjusted. In social network contexts, a disagreement in opinion between individuals may result in these individuals ceasing interaction, thus removing an edge - this corresponds to an adaptive voter model, and could result in the formation of echo-chambers over the network centered on the zealots. The individuals within a social network are often influenced by effects not captured in the network, meaning some may flip opinion spontaneously due to extraneous factors - to capture this, a noisy voter model ought to be considered. A review of many of the variations of the voter model is found in Ref.\ \cite{redner2019reality}. In cases where de-conversion from the weak opinion results in reluctance to accept it again, an $SIR$-like model \cite{kermack1927contribution} may be preferable to the voter model, and has the benefit of also capturing an epidemiological version of this problem where zealots correspond to a source of infection which can go on to spread through the population. In addition to this, social networks described by different varieties of graphs may better suit specific case-studies. For example, in cases where interactions are based on physical proximity we ought to employ a random geometric graph \cite{penrose2003random}, in which nodes are embedded in $\mathbb{R}^2$, and edges exist between within some distance $d$. In this scenario, we predict that the long-term distribution of the weak opinion may again be estimated by accounting for the distribution of small components and giant components separately, as in Eqn.\ \ref{FULL}. 


\textbf{Acknowledgements:} Thomas Tunstall acknowledges support by EPSRC DTP and Syngenta Crop Protection, and thanks Wolfram M{\"o}bius as well as Tim Rogers for guidance and and discussions on the work and feedback on a draft of this manuscript. Furthermore, he thanks Mauro Mobilia for comments and feedback during on an early version of this manuscript. 

\bibliography{bibliography}

\begin{thebibliography}{39}%
\makeatletter
\providecommand \@ifxundefined [1]{%
 \@ifx{#1\undefined}
}%
\providecommand \@ifnum [1]{%
 \ifnum #1\expandafter \@firstoftwo
 \else \expandafter \@secondoftwo
 \fi
}%
\providecommand \@ifx [1]{%
 \ifx #1\expandafter \@firstoftwo
 \else \expandafter \@secondoftwo
 \fi
}%
\providecommand \natexlab [1]{#1}%
\providecommand \enquote  [1]{``#1''}%
\providecommand \bibnamefont  [1]{#1}%
\providecommand \bibfnamefont [1]{#1}%
\providecommand \citenamefont [1]{#1}%
\providecommand \href@noop [0]{\@secondoftwo}%
\providecommand \href [0]{\begingroup \@sanitize@url \@href}%
\providecommand \@href[1]{\@@startlink{#1}\@@href}%
\providecommand \@@href[1]{\endgroup#1\@@endlink}%
\providecommand \@sanitize@url [0]{\catcode `\\12\catcode `\$12\catcode `\&12\catcode `\#12\catcode `\^12\catcode `\_12\catcode `\%12\relax}%
\providecommand \@@startlink[1]{}%
\providecommand \@@endlink[0]{}%
\providecommand \url  [0]{\begingroup\@sanitize@url \@url }%
\providecommand \@url [1]{\endgroup\@href {#1}{\urlprefix }}%
\providecommand \urlprefix  [0]{URL }%
\providecommand \Eprint [0]{\href }%
\providecommand \doibase [0]{http://dx.doi.org/}%
\providecommand \selectlanguage [0]{\@gobble}%
\providecommand \bibinfo  [0]{\@secondoftwo}%
\providecommand \bibfield  [0]{\@secondoftwo}%
\providecommand \translation [1]{[#1]}%
\providecommand \BibitemOpen [0]{}%
\providecommand \bibitemStop [0]{}%
\providecommand \bibitemNoStop [0]{.\EOS\space}%
\providecommand \EOS [0]{\spacefactor3000\relax}%
\providecommand \BibitemShut  [1]{\csname bibitem#1\endcsname}%
\let\auto@bib@innerbib\@empty
\bibitem [{\citenamefont {Clifford}\ and\ \citenamefont {Sudbury}(1973)}]{clifford1973model}%
  \BibitemOpen
  \bibfield  {author} {\bibinfo {author} {\bibfnamefont {P.}~\bibnamefont {Clifford}}\ and\ \bibinfo {author} {\bibfnamefont {A.}~\bibnamefont {Sudbury}},\ }\href@noop {} {\bibfield  {journal} {\bibinfo  {journal} {Biometrika}\ }\textbf {\bibinfo {volume} {60}},\ \bibinfo {pages} {581} (\bibinfo {year} {1973})}\BibitemShut {NoStop}%
\bibitem [{\citenamefont {Holley}\ and\ \citenamefont {Liggett}(1975)}]{holley1975ergodic}%
  \BibitemOpen
  \bibfield  {author} {\bibinfo {author} {\bibfnamefont {R.~A.}\ \bibnamefont {Holley}}\ and\ \bibinfo {author} {\bibfnamefont {T.~M.}\ \bibnamefont {Liggett}},\ }\href@noop {} {\bibfield  {journal} {\bibinfo  {journal} {The annals of probability}\ ,\ \bibinfo {pages} {643}} (\bibinfo {year} {1975})}\BibitemShut {NoStop}%
\bibitem [{\citenamefont {Krapivsky}(1992)}]{krapivsky1992kinetics}%
  \BibitemOpen
  \bibfield  {author} {\bibinfo {author} {\bibfnamefont {P.~L.}\ \bibnamefont {Krapivsky}},\ }\href@noop {} {\bibfield  {journal} {\bibinfo  {journal} {Physical Review A}\ }\textbf {\bibinfo {volume} {45}},\ \bibinfo {pages} {1067} (\bibinfo {year} {1992})}\BibitemShut {NoStop}%
\bibitem [{\citenamefont {Frachebourg}\ and\ \citenamefont {Krapivsky}(1996)}]{frachebourg1996exact}%
  \BibitemOpen
  \bibfield  {author} {\bibinfo {author} {\bibfnamefont {L.}~\bibnamefont {Frachebourg}}\ and\ \bibinfo {author} {\bibfnamefont {P.~L.}\ \bibnamefont {Krapivsky}},\ }\href@noop {} {\bibfield  {journal} {\bibinfo  {journal} {Physical Review E}\ }\textbf {\bibinfo {volume} {53}},\ \bibinfo {pages} {R3009} (\bibinfo {year} {1996})}\BibitemShut {NoStop}%
\bibitem [{\citenamefont {Castellano}\ \emph {et~al.}(2003)\citenamefont {Castellano}, \citenamefont {Vilone},\ and\ \citenamefont {Vespignani}}]{castellano2003incomplete}%
  \BibitemOpen
  \bibfield  {author} {\bibinfo {author} {\bibfnamefont {C.}~\bibnamefont {Castellano}}, \bibinfo {author} {\bibfnamefont {D.}~\bibnamefont {Vilone}}, \ and\ \bibinfo {author} {\bibfnamefont {A.}~\bibnamefont {Vespignani}},\ }\href@noop {} {\bibfield  {journal} {\bibinfo  {journal} {Europhysics Letters}\ }\textbf {\bibinfo {volume} {63}},\ \bibinfo {pages} {153} (\bibinfo {year} {2003})}\BibitemShut {NoStop}%
\bibitem [{\citenamefont {Sood}\ and\ \citenamefont {Redner}(2005)}]{sood2005voter}%
  \BibitemOpen
  \bibfield  {author} {\bibinfo {author} {\bibfnamefont {V.}~\bibnamefont {Sood}}\ and\ \bibinfo {author} {\bibfnamefont {S.}~\bibnamefont {Redner}},\ }\href@noop {} {\bibfield  {journal} {\bibinfo  {journal} {Physical review letters}\ }\textbf {\bibinfo {volume} {94}},\ \bibinfo {pages} {178701} (\bibinfo {year} {2005})}\BibitemShut {NoStop}%
\bibitem [{\citenamefont {Castellano}\ \emph {et~al.}(2009)\citenamefont {Castellano}, \citenamefont {Fortunato},\ and\ \citenamefont {Loreto}}]{castellano2009statistical}%
  \BibitemOpen
  \bibfield  {author} {\bibinfo {author} {\bibfnamefont {C.}~\bibnamefont {Castellano}}, \bibinfo {author} {\bibfnamefont {S.}~\bibnamefont {Fortunato}}, \ and\ \bibinfo {author} {\bibfnamefont {V.}~\bibnamefont {Loreto}},\ }\href@noop {} {\bibfield  {journal} {\bibinfo  {journal} {Reviews of modern physics}\ }\textbf {\bibinfo {volume} {81}},\ \bibinfo {pages} {591} (\bibinfo {year} {2009})}\BibitemShut {NoStop}%
\bibitem [{\citenamefont {Yildiz}\ \emph {et~al.}(2010)\citenamefont {Yildiz}, \citenamefont {Pagliari}, \citenamefont {Ozdaglar},\ and\ \citenamefont {Scaglione}}]{yildiz2010voting}%
  \BibitemOpen
  \bibfield  {author} {\bibinfo {author} {\bibfnamefont {M.~E.}\ \bibnamefont {Yildiz}}, \bibinfo {author} {\bibfnamefont {R.}~\bibnamefont {Pagliari}}, \bibinfo {author} {\bibfnamefont {A.}~\bibnamefont {Ozdaglar}}, \ and\ \bibinfo {author} {\bibfnamefont {A.}~\bibnamefont {Scaglione}},\ }in\ \href@noop {} {\emph {\bibinfo {booktitle} {2010 information theory and applications workshop (ITA)}}}\ (\bibinfo {organization} {IEEE},\ \bibinfo {year} {2010})\ pp.\ \bibinfo {pages} {1--7}\BibitemShut {NoStop}%
\bibitem [{\citenamefont {Redner}(2019)}]{redner2019reality}%
  \BibitemOpen
  \bibfield  {author} {\bibinfo {author} {\bibfnamefont {S.}~\bibnamefont {Redner}},\ }\href@noop {} {\bibfield  {journal} {\bibinfo  {journal} {Comptes Rendus Physique}\ }\textbf {\bibinfo {volume} {20}},\ \bibinfo {pages} {275} (\bibinfo {year} {2019})}\BibitemShut {NoStop}%
\bibitem [{\citenamefont {Tang}\ \emph {et~al.}(2015)\citenamefont {Tang}, \citenamefont {Yan}, \citenamefont {Pei},\ and\ \citenamefont {Zheng}}]{tang2015evolutionary}%
  \BibitemOpen
  \bibfield  {author} {\bibinfo {author} {\bibfnamefont {S.}~\bibnamefont {Tang}}, \bibinfo {author} {\bibfnamefont {S.}~\bibnamefont {Yan}}, \bibinfo {author} {\bibfnamefont {S.}~\bibnamefont {Pei}}, \ and\ \bibinfo {author} {\bibfnamefont {Z.}~\bibnamefont {Zheng}},\ }\href@noop {} {\bibfield  {journal} {\bibinfo  {journal} {Journal of the Korean Physical Society}\ }\textbf {\bibinfo {volume} {66}},\ \bibinfo {pages} {1783} (\bibinfo {year} {2015})}\BibitemShut {NoStop}%
\bibitem [{\citenamefont {Mukhopadhyay}\ \emph {et~al.}(2020)\citenamefont {Mukhopadhyay}, \citenamefont {Mazumdar},\ and\ \citenamefont {Roy}}]{mukhopadhyay2020voter}%
  \BibitemOpen
  \bibfield  {author} {\bibinfo {author} {\bibfnamefont {A.}~\bibnamefont {Mukhopadhyay}}, \bibinfo {author} {\bibfnamefont {R.~R.}\ \bibnamefont {Mazumdar}}, \ and\ \bibinfo {author} {\bibfnamefont {R.}~\bibnamefont {Roy}},\ }\href@noop {} {\bibfield  {journal} {\bibinfo  {journal} {Journal of Statistical Physics}\ }\textbf {\bibinfo {volume} {181}},\ \bibinfo {pages} {1239} (\bibinfo {year} {2020})}\BibitemShut {NoStop}%
\bibitem [{\citenamefont {Antal}\ \emph {et~al.}(2006)\citenamefont {Antal}, \citenamefont {Redner},\ and\ \citenamefont {Sood}}]{antal2006evolutionary}%
  \BibitemOpen
  \bibfield  {author} {\bibinfo {author} {\bibfnamefont {T.}~\bibnamefont {Antal}}, \bibinfo {author} {\bibfnamefont {S.}~\bibnamefont {Redner}}, \ and\ \bibinfo {author} {\bibfnamefont {V.}~\bibnamefont {Sood}},\ }\href@noop {} {\bibfield  {journal} {\bibinfo  {journal} {Physical review letters}\ }\textbf {\bibinfo {volume} {96}},\ \bibinfo {pages} {188104} (\bibinfo {year} {2006})}\BibitemShut {NoStop}%
\bibitem [{\citenamefont {Sood}\ \emph {et~al.}(2008)\citenamefont {Sood}, \citenamefont {Antal},\ and\ \citenamefont {Redner}}]{sood2008voter}%
  \BibitemOpen
  \bibfield  {author} {\bibinfo {author} {\bibfnamefont {V.}~\bibnamefont {Sood}}, \bibinfo {author} {\bibfnamefont {T.}~\bibnamefont {Antal}}, \ and\ \bibinfo {author} {\bibfnamefont {S.}~\bibnamefont {Redner}},\ }\href@noop {} {\bibfield  {journal} {\bibinfo  {journal} {Physical Review E}\ }\textbf {\bibinfo {volume} {77}},\ \bibinfo {pages} {041121} (\bibinfo {year} {2008})}\BibitemShut {NoStop}%
\bibitem [{\citenamefont {Mobilia}(2003)}]{mobilia2003does}%
  \BibitemOpen
  \bibfield  {author} {\bibinfo {author} {\bibfnamefont {M.}~\bibnamefont {Mobilia}},\ }\href@noop {} {\bibfield  {journal} {\bibinfo  {journal} {Physical review letters}\ }\textbf {\bibinfo {volume} {91}},\ \bibinfo {pages} {028701} (\bibinfo {year} {2003})}\BibitemShut {NoStop}%
\bibitem [{\citenamefont {Mobilia}\ and\ \citenamefont {Georgiev}(2005)}]{mobilia2005voting}%
  \BibitemOpen
  \bibfield  {author} {\bibinfo {author} {\bibfnamefont {M.}~\bibnamefont {Mobilia}}\ and\ \bibinfo {author} {\bibfnamefont {I.~T.}\ \bibnamefont {Georgiev}},\ }\href@noop {} {\bibfield  {journal} {\bibinfo  {journal} {Physical Review E}\ }\textbf {\bibinfo {volume} {71}},\ \bibinfo {pages} {046102} (\bibinfo {year} {2005})}\BibitemShut {NoStop}%
\bibitem [{\citenamefont {Caligiuri}\ and\ \citenamefont {Galla}(2023)}]{caligiuri2023noisy}%
  \BibitemOpen
  \bibfield  {author} {\bibinfo {author} {\bibfnamefont {A.}~\bibnamefont {Caligiuri}}\ and\ \bibinfo {author} {\bibfnamefont {T.}~\bibnamefont {Galla}},\ }\href@noop {} {\bibfield  {journal} {\bibinfo  {journal} {Physical Review E}\ }\textbf {\bibinfo {volume} {108}},\ \bibinfo {pages} {044301} (\bibinfo {year} {2023})}\BibitemShut {NoStop}%
\bibitem [{\citenamefont {Mobilia}\ \emph {et~al.}(2007)\citenamefont {Mobilia}, \citenamefont {Petersen},\ and\ \citenamefont {Redner}}]{mobilia2007role}%
  \BibitemOpen
  \bibfield  {author} {\bibinfo {author} {\bibfnamefont {M.}~\bibnamefont {Mobilia}}, \bibinfo {author} {\bibfnamefont {A.}~\bibnamefont {Petersen}}, \ and\ \bibinfo {author} {\bibfnamefont {S.}~\bibnamefont {Redner}},\ }\href@noop {} {\bibfield  {journal} {\bibinfo  {journal} {Journal of Statistical Mechanics: Theory and Experiment}\ }\textbf {\bibinfo {volume} {2007}},\ \bibinfo {pages} {P08029} (\bibinfo {year} {2007})}\BibitemShut {NoStop}%
\bibitem [{\citenamefont {Mobilia}(2015)}]{mobilia2015nonlinear}%
  \BibitemOpen
  \bibfield  {author} {\bibinfo {author} {\bibfnamefont {M.}~\bibnamefont {Mobilia}},\ }\href@noop {} {\bibfield  {journal} {\bibinfo  {journal} {Physical Review E}\ }\textbf {\bibinfo {volume} {92}},\ \bibinfo {pages} {012803} (\bibinfo {year} {2015})}\BibitemShut {NoStop}%
\bibitem [{\citenamefont {Galam}\ and\ \citenamefont {Jacobs}(2007)}]{galam2007role}%
  \BibitemOpen
  \bibfield  {author} {\bibinfo {author} {\bibfnamefont {S.}~\bibnamefont {Galam}}\ and\ \bibinfo {author} {\bibfnamefont {F.}~\bibnamefont {Jacobs}},\ }\href@noop {} {\bibfield  {journal} {\bibinfo  {journal} {Physica A: Statistical Mechanics and its Applications}\ }\textbf {\bibinfo {volume} {381}},\ \bibinfo {pages} {366} (\bibinfo {year} {2007})}\BibitemShut {NoStop}%
\bibitem [{\citenamefont {Vendeville}\ \emph {et~al.}(2022)\citenamefont {Vendeville}, \citenamefont {Guedj},\ and\ \citenamefont {Zhou}}]{vendeville2022towards}%
  \BibitemOpen
  \bibfield  {author} {\bibinfo {author} {\bibfnamefont {A.}~\bibnamefont {Vendeville}}, \bibinfo {author} {\bibfnamefont {B.}~\bibnamefont {Guedj}}, \ and\ \bibinfo {author} {\bibfnamefont {S.}~\bibnamefont {Zhou}},\ }in\ \href@noop {} {\emph {\bibinfo {booktitle} {Complex Networks \& Their Applications X: Volume 2, Proceedings of the Tenth International Conference on Complex Networks and Their Applications COMPLEX NETWORKS 2021 10}}}\ (\bibinfo {organization} {Springer},\ \bibinfo {year} {2022})\ pp.\ \bibinfo {pages} {341--352}\BibitemShut {NoStop}%
\bibitem [{\citenamefont {Moeinifar}\ and\ \citenamefont {G{\"u}nd{\"u}{\c{c}}}(2021)}]{moeinifar2021zealots}%
  \BibitemOpen
  \bibfield  {author} {\bibinfo {author} {\bibfnamefont {V.}~\bibnamefont {Moeinifar}}\ and\ \bibinfo {author} {\bibfnamefont {S.}~\bibnamefont {G{\"u}nd{\"u}{\c{c}}}},\ }\href@noop {} {\bibfield  {journal} {\bibinfo  {journal} {Mathematical Modeling and Computing}\ }\textbf {\bibinfo {volume} {8}} (\bibinfo {year} {2021})}\BibitemShut {NoStop}%
\bibitem [{\citenamefont {Czaplicka}\ \emph {et~al.}(2022)\citenamefont {Czaplicka}, \citenamefont {Charalambous}, \citenamefont {Toral},\ and\ \citenamefont {San~Miguel}}]{czaplicka2022biased}%
  \BibitemOpen
  \bibfield  {author} {\bibinfo {author} {\bibfnamefont {A.}~\bibnamefont {Czaplicka}}, \bibinfo {author} {\bibfnamefont {C.}~\bibnamefont {Charalambous}}, \bibinfo {author} {\bibfnamefont {R.}~\bibnamefont {Toral}}, \ and\ \bibinfo {author} {\bibfnamefont {M.}~\bibnamefont {San~Miguel}},\ }\href@noop {} {\bibfield  {journal} {\bibinfo  {journal} {Chaos, Solitons \& Fractals}\ }\textbf {\bibinfo {volume} {161}},\ \bibinfo {pages} {112363} (\bibinfo {year} {2022})}\BibitemShut {NoStop}%
\bibitem [{\citenamefont {Braha}\ and\ \citenamefont {De~Aguiar}(2017)}]{braha2017voting}%
  \BibitemOpen
  \bibfield  {author} {\bibinfo {author} {\bibfnamefont {D.}~\bibnamefont {Braha}}\ and\ \bibinfo {author} {\bibfnamefont {M.~A.}\ \bibnamefont {De~Aguiar}},\ }\href@noop {} {\bibfield  {journal} {\bibinfo  {journal} {PloS one}\ }\textbf {\bibinfo {volume} {12}},\ \bibinfo {pages} {e0177970} (\bibinfo {year} {2017})}\BibitemShut {NoStop}%
\bibitem [{\citenamefont {M{\"u}ller}\ \emph {et~al.}(2022)\citenamefont {M{\"u}ller}, \citenamefont {Tellier},\ and\ \citenamefont {Kurschilgen}}]{muller2022echo}%
  \BibitemOpen
  \bibfield  {author} {\bibinfo {author} {\bibfnamefont {J.}~\bibnamefont {M{\"u}ller}}, \bibinfo {author} {\bibfnamefont {A.}~\bibnamefont {Tellier}}, \ and\ \bibinfo {author} {\bibfnamefont {M.}~\bibnamefont {Kurschilgen}},\ }\href@noop {} {\bibfield  {journal} {\bibinfo  {journal} {Royal Society Open Science}\ }\textbf {\bibinfo {volume} {9}},\ \bibinfo {pages} {220367} (\bibinfo {year} {2022})}\BibitemShut {NoStop}%
\bibitem [{\citenamefont {Barrat}\ \emph {et~al.}(2008)\citenamefont {Barrat}, \citenamefont {Barthelemy},\ and\ \citenamefont {Vespignani}}]{barrat2008dynamical}%
  \BibitemOpen
  \bibfield  {author} {\bibinfo {author} {\bibfnamefont {A.}~\bibnamefont {Barrat}}, \bibinfo {author} {\bibfnamefont {M.}~\bibnamefont {Barthelemy}}, \ and\ \bibinfo {author} {\bibfnamefont {A.}~\bibnamefont {Vespignani}},\ }\href@noop {} {\emph {\bibinfo {title} {Dynamical processes on complex networks}}}\ (\bibinfo  {publisher} {Cambridge university press},\ \bibinfo {year} {2008})\BibitemShut {NoStop}%
\bibitem [{\citenamefont {Newman}(2001)}]{newman2001structure}%
  \BibitemOpen
  \bibfield  {author} {\bibinfo {author} {\bibfnamefont {M.~E.}\ \bibnamefont {Newman}},\ }\href@noop {} {\bibfield  {journal} {\bibinfo  {journal} {Proceedings of the national academy of sciences}\ }\textbf {\bibinfo {volume} {98}},\ \bibinfo {pages} {404} (\bibinfo {year} {2001})}\BibitemShut {NoStop}%
\bibitem [{\citenamefont {Erd{\H o}s}\ and\ \citenamefont {R{\'e}nyi}(1959)}]{erdds1959random}%
  \BibitemOpen
  \bibfield  {author} {\bibinfo {author} {\bibfnamefont {P.}~\bibnamefont {Erd{\H o}s}}\ and\ \bibinfo {author} {\bibfnamefont {A.}~\bibnamefont {R{\'e}nyi}},\ }\href@noop {} {\bibfield  {journal} {\bibinfo  {journal} {Publ. math. debrecen}\ }\textbf {\bibinfo {volume} {6}},\ \bibinfo {pages} {18} (\bibinfo {year} {1959})}\BibitemShut {NoStop}%
\bibitem [{\citenamefont {Moran}(1958)}]{moran1958random}%
  \BibitemOpen
  \bibfield  {author} {\bibinfo {author} {\bibfnamefont {P.~A.~P.}\ \bibnamefont {Moran}},\ }in\ \href@noop {} {\emph {\bibinfo {booktitle} {Mathematical proceedings of the cambridge philosophical society}}},\ Vol.~\bibinfo {volume} {54}\ (\bibinfo {organization} {Cambridge University Press},\ \bibinfo {year} {1958})\ pp.\ \bibinfo {pages} {60--71}\BibitemShut {NoStop}%
\bibitem [{\citenamefont {Nowak}(2006)}]{nowak2006evolutionary}%
  \BibitemOpen
  \bibfield  {author} {\bibinfo {author} {\bibfnamefont {M.~A.}\ \bibnamefont {Nowak}},\ }\href@noop {} {\emph {\bibinfo {title} {Evolutionary dynamics: exploring the equations of life}}}\ (\bibinfo  {publisher} {Harvard university press},\ \bibinfo {year} {2006})\BibitemShut {NoStop}%
\bibitem [{\citenamefont {Masuda}(2012)}]{masuda2012evolution}%
  \BibitemOpen
  \bibfield  {author} {\bibinfo {author} {\bibfnamefont {N.}~\bibnamefont {Masuda}},\ }\href@noop {} {\bibfield  {journal} {\bibinfo  {journal} {Scientific reports}\ }\textbf {\bibinfo {volume} {2}},\ \bibinfo {pages} {646} (\bibinfo {year} {2012})}\BibitemShut {NoStop}%
\bibitem [{\citenamefont {Cardillo}\ and\ \citenamefont {Masuda}(2020)}]{cardillo2020critical}%
  \BibitemOpen
  \bibfield  {author} {\bibinfo {author} {\bibfnamefont {A.}~\bibnamefont {Cardillo}}\ and\ \bibinfo {author} {\bibfnamefont {N.}~\bibnamefont {Masuda}},\ }\href@noop {} {\bibfield  {journal} {\bibinfo  {journal} {Physical Review Research}\ }\textbf {\bibinfo {volume} {2}},\ \bibinfo {pages} {023305} (\bibinfo {year} {2020})}\BibitemShut {NoStop}%
\bibitem [{\citenamefont {Andersson}\ and\ \citenamefont {Levin}(1999)}]{andersson_biological_1999}%
  \BibitemOpen
  \bibfield  {author} {\bibinfo {author} {\bibfnamefont {D.~I.}\ \bibnamefont {Andersson}}\ and\ \bibinfo {author} {\bibfnamefont {B.~R.}\ \bibnamefont {Levin}},\ }\href@noop {} {\bibfield  {journal} {\bibinfo  {journal} {Current Opinion in Microbiology}\ }\textbf {\bibinfo {volume} {2}},\ \bibinfo {pages} {489} (\bibinfo {year} {1999})}\BibitemShut {NoStop}%
\bibitem [{\citenamefont {Kliot}\ and\ \citenamefont {Ghanim}(2012)}]{kliot2012fitness}%
  \BibitemOpen
  \bibfield  {author} {\bibinfo {author} {\bibfnamefont {A.}~\bibnamefont {Kliot}}\ and\ \bibinfo {author} {\bibfnamefont {M.}~\bibnamefont {Ghanim}},\ }\href@noop {} {\bibfield  {journal} {\bibinfo  {journal} {Pest management science}\ }\textbf {\bibinfo {volume} {68}},\ \bibinfo {pages} {1431} (\bibinfo {year} {2012})}\BibitemShut {NoStop}%
\bibitem [{\citenamefont {Erd{\H{o}}s}\ \emph {et~al.}(1960)\citenamefont {Erd{\H{o}}s}, \citenamefont {R{\'e}nyi} \emph {et~al.}}]{erdHos1960evolution}%
  \BibitemOpen
  \bibfield  {author} {\bibinfo {author} {\bibfnamefont {P.}~\bibnamefont {Erd{\H{o}}s}}, \bibinfo {author} {\bibfnamefont {A.}~\bibnamefont {R{\'e}nyi}},  \emph {et~al.},\ }\href@noop {} {\bibfield  {journal} {\bibinfo  {journal} {Publ. math. inst. hung. acad. sci}\ }\textbf {\bibinfo {volume} {5}},\ \bibinfo {pages} {17} (\bibinfo {year} {1960})}\BibitemShut {NoStop}%
\bibitem [{\citenamefont {Bollob{\'a}s}(2001)}]{bollobas_random_2001}%
  \BibitemOpen
  \bibfield  {author} {\bibinfo {author} {\bibfnamefont {B.}~\bibnamefont {Bollob{\'a}s}},\ }\href {\doibase 10.1017/CBO9780511814068} {\emph {\bibinfo {title} {Random {Graphs}}}},\ \bibinfo {edition} {2nd}\ ed.,\ Cambridge {Studies} in {Advanced} {Mathematics}\ (\bibinfo  {publisher} {Cambridge University Press},\ \bibinfo {address} {Cambridge},\ \bibinfo {year} {2001})\BibitemShut {NoStop}%
\bibitem [{Sup()}]{Supplement}%
  \BibitemOpen
  \href@noop {} {}\bibinfo {note} {See Supplemental Material at [URL will be inserted by publisher] for (1) a derivation of the mean-field Fokker-Planck equation; (2) a derivation of the stability of the deterministic equilibria of the Fokker-Planck equation. The Supplemental Material also contains Refs. \cite{mobilia2007role}.}\BibitemShut {Stop}%
\bibitem [{\citenamefont {Tishby}\ \emph {et~al.}(2018)\citenamefont {Tishby}, \citenamefont {Biham}, \citenamefont {Katzav},\ and\ \citenamefont {K{\"u}hn}}]{tishby2018revealing}%
  \BibitemOpen
  \bibfield  {author} {\bibinfo {author} {\bibfnamefont {I.}~\bibnamefont {Tishby}}, \bibinfo {author} {\bibfnamefont {O.}~\bibnamefont {Biham}}, \bibinfo {author} {\bibfnamefont {E.}~\bibnamefont {Katzav}}, \ and\ \bibinfo {author} {\bibfnamefont {R.}~\bibnamefont {K{\"u}hn}},\ }\href@noop {} {\bibfield  {journal} {\bibinfo  {journal} {Physical Review E}\ }\textbf {\bibinfo {volume} {97}},\ \bibinfo {pages} {042318} (\bibinfo {year} {2018})}\BibitemShut {NoStop}%
\bibitem [{\citenamefont {Kermack}\ and\ \citenamefont {McKendrick}(1927)}]{kermack1927contribution}%
  \BibitemOpen
  \bibfield  {author} {\bibinfo {author} {\bibfnamefont {W.~O.}\ \bibnamefont {Kermack}}\ and\ \bibinfo {author} {\bibfnamefont {A.~G.}\ \bibnamefont {McKendrick}},\ }\href@noop {} {\bibfield  {journal} {\bibinfo  {journal} {Proceedings of the royal society of london. Series A, Containing papers of a mathematical and physical character}\ }\textbf {\bibinfo {volume} {115}},\ \bibinfo {pages} {700} (\bibinfo {year} {1927})}\BibitemShut {NoStop}%
\bibitem [{\citenamefont {Penrose}(2003)}]{penrose2003random}%
  \BibitemOpen
  \bibfield  {author} {\bibinfo {author} {\bibfnamefont {M.}~\bibnamefont {Penrose}},\ }\href@noop {} {\emph {\bibinfo {title} {Random geometric graphs}}},\ Vol.~\bibinfo {volume} {5}\ (\bibinfo  {publisher} {OUP Oxford},\ \bibinfo {year} {2003})\BibitemShut {NoStop}%
\end{thebibliography}%


\begin{thebibliography}{1}%
\makeatletter
\providecommand \@ifxundefined [1]{%
 \@ifx{#1\undefined}
}%
\providecommand \@ifnum [1]{%
 \ifnum #1\expandafter \@firstoftwo
 \else \expandafter \@secondoftwo
 \fi
}%
\providecommand \@ifx [1]{%
 \ifx #1\expandafter \@firstoftwo
 \else \expandafter \@secondoftwo
 \fi
}%
\providecommand \natexlab [1]{#1}%
\providecommand \enquote  [1]{``#1''}%
\providecommand \bibnamefont  [1]{#1}%
\providecommand \bibfnamefont [1]{#1}%
\providecommand \citenamefont [1]{#1}%
\providecommand \href@noop [0]{\@secondoftwo}%
\providecommand \href [0]{\begingroup \@sanitize@url \@href}%
\providecommand \@href[1]{\@@startlink{#1}\@@href}%
\providecommand \@@href[1]{\endgroup#1\@@endlink}%
\providecommand \@sanitize@url [0]{\catcode `\\12\catcode `\$12\catcode `\&12\catcode `\#12\catcode `\^12\catcode `\_12\catcode `\%12\relax}%
\providecommand \@@startlink[1]{}%
\providecommand \@@endlink[0]{}%
\providecommand \url  [0]{\begingroup\@sanitize@url \@url }%
\providecommand \@url [1]{\endgroup\@href {#1}{\urlprefix }}%
\providecommand \urlprefix  [0]{URL }%
\providecommand \Eprint [0]{\href }%
\providecommand \doibase [0]{http://dx.doi.org/}%
\providecommand \selectlanguage [0]{\@gobble}%
\providecommand \bibinfo  [0]{\@secondoftwo}%
\providecommand \bibfield  [0]{\@secondoftwo}%
\providecommand \translation [1]{[#1]}%
\providecommand \BibitemOpen [0]{}%
\providecommand \bibitemStop [0]{}%
\providecommand \bibitemNoStop [0]{.\EOS\space}%
\providecommand \EOS [0]{\spacefactor3000\relax}%
\providecommand \BibitemShut  [1]{\csname bibitem#1\endcsname}%
\let\auto@bib@innerbib\@empty
\bibitem [{\citenamefont {Mobilia}, \citenamefont {Petersen},\ and\ \citenamefont {Redner}(2007)}]{mobilia2007role}%
  \BibitemOpen
  \bibfield  {author} {\bibinfo {author} {\bibfnamefont {M.}~\bibnamefont {Mobilia}}, \bibinfo {author} {\bibfnamefont {A.}~\bibnamefont {Petersen}}, \ and\ \bibinfo {author} {\bibfnamefont {S.}~\bibnamefont {Redner}},\ }\href@noop {} {\bibfield  {journal} {\bibinfo  {journal} {Journal of Statistical Mechanics: Theory and Experiment}\ }\textbf {\bibinfo {volume} {2007}},\ \bibinfo {pages} {P08029} (\bibinfo {year} {2007})}\BibitemShut {NoStop}%
\end{thebibliography}%

\end{document}


\title{Supplementary Material to ``How Social Network Structure Impacts the Ability of Zealots to Promote Weak Opinions''}

\author[1,2,3]{Thomas Tunstall}

\affil[1]{Living Systems Institute, Faculty of Health and Life Sciences, University of Exeter}
\affil[2]{Physics and Astronomy, Faculty of Environment, Science and Economy, University of Exeter}
\affil[3]{Mathematics and Statistics, Faculty of Environment, Science and Economy, University of Exeter}

\maketitle

\tableofcontents

\section{Derivation of the Fokker-Planck equation}

We consider a complete graph of $N$ nodes, where $Z$ are zealots which promote a `weak' opinion which is accepted by a neighbour with a probability $F:F<1$ relative to a strong opinion. The remaining $N_{\text{free}}$ nodes may change their opinion, and for a given state of the system there are $N_S$ free nodes subscribed to the strong opinion and $N_W$ free nodes subscribed to the weak opinion (such that $N_S + N_W = N_{\text{free}}$).

The derivation for the steady-state distribution of the weaker opinion across the complete graph of size $N$ can be obtained by employing an argument similar to that presented in \cite{mobilia2007role}, except for the implementation of only one variety of zealot, and applying a relative fitness factor $F$ which lowers the probability that a weaker opinion neighbour is selected. The probability of there being $N_W$ free nodes which subscribe to the weaker opinion at time $t$ is denoted by $P(N_W,t)$: the corresponding master equation is:

\begin{equation}\label{MasterEquation}
    \begin{split}
        \frac{\partial P(N_W,t)}{\partial t} &= \sum_{\delta=\pm 1} P(N_W+\delta,t) W(N_W + \delta\to N_W)\\
        &-\sum_{\delta=\pm 1} P(N_W,t) W(N_W \to N_W+\delta).
    \end{split}
\end{equation}

The first term corresponds to the rate at which the the $N_W$ state is entered, and the second corresponds to the rate at which the $N_W$ state is exited. The transition rates, $W$, are given by

\begin{equation}\label{TransitionRates}
    \begin{split}
        \delta t W(N_W\to N_W-1) &= \frac{N_W}{N_{\text{free}}} \times \frac{N_S}{N_S + F(N_W+Z)},\\
        \delta t W(N_W\to N_W+1) &= \frac{N_S}{N_{\text{free}}} \times \frac{F(N_R+Z)}{N_S + F(N_W+Z)}.
    \end{split}
\end{equation}

$\delta t W(N_W\to N_W-1)$ corresponds to the probability of the state decreasing from $N_W$ to $N_W-1$ during a single update step, and corresponds to the product of the independent probabilities that a weaker free node is selected to update its opinion (first term), and then that it updates to the stronger opinion (second term). Correspondingly, $\delta t W(N_W\to N_W+1)$ corresponds to the probability of the state increasing from $N_W$ to $N_W+1$ during a single update step, and corresponds to the product of the independent probabilities that a stronger free node is selected to update its opinion, and then that it updates to the weaker opinion. As in Ref.\ \cite{mobilia2007role}, we shall take $\delta t$ to be $N^{-1}$ to represent that on average all nodes update their opinion per unit time. We shall also derive an analytical solution in the continuum limit $N\to\infty$, making the change in variables $z = Z/N$, $n = N_W/N$, $n_S = N_S/N$, $f = N_{\text{free}}/N = 1-z$ to correspond to proportions rather than number of nodes. Given these choices and recognising $n_S = f - n$, we follow the steps laid out in \cite{mobilia2007role} to obtain the Fokker-Planck equation:

\begin{equation}\label{FP}
    \begin{split}
        \frac{\partial P(n,t)}{\partial t} &= -\frac{\partial}{\partial n}\left[\alpha(n) P(n,t)\right] \\
        &+ \frac{1}{2} \frac{\partial^2}{\partial n^2}\left[\beta(n)P(n,t)\right],
    \end{split}
\end{equation}

in which

\begin{equation}\label{alphabeta}
    \begin{split}
        \alpha(n;N,F,z)&= \frac{1-z-n}{1-z}\frac{F(n+z) -n}{1-z-n+F(n+z)},\\
        \beta(n;N,F,z)&=  \frac{1-z-n}{(1-z)N}\frac{F(n+z) + n}{1-z-n+F(n+z)}.
    \end{split}
\end{equation}

As in Ref.\ \cite{mobilia2007role}, the probability of a proportion $n$ of the free nodes adopting the weak opinion in the long term is:

\begin{equation*}
     P(n;N,F,z) = \mathcal{Z} \frac{\exp{2 \int_0^n dn' \alpha(n')/\beta(n')}}{\beta(n)},
\end{equation*}

where $\mathcal{Z}$ is a normalisation factor.

In order to simulate such a system, it is necessary to introduce a single zealot of the strong variety in order to ensure that the absorbing state (in which all nodes adopt the weak opinion) can never be entered. This can be implemented by defining $Z_S=1$, and updating the transition rates (Eqn.\ \ref{TransitionRates}):

\begin{equation}\label{TransitionRates_New}
    \begin{split}
        \delta t W(N_W\to N_W-1) &= \frac{N_W}{N_{\text{free}}} \times \frac{N_S + Z_S}{N_S + Z_S + F(N_W+Z)},\\
        \delta t W(N_W\to N_W+1) &= \frac{N_S}{N_{\text{free}}} \times \frac{F(N_R+Z)}{N_S + Z_S + F(N_W+Z)}.
    \end{split}
\end{equation}

Making the change in variables again, (setting $z_S = Z_S/N$), we get an identical Fokker-Planck equation as Eqn.\ \ref{FP}, in which $\alpha(n;N,F,z,z_S)$ and $\beta(n;N,F,z_S)$ take the form: 

\begin{equation}\label{alphabeta_new}
    \begin{split}
        \alpha(n;N,F,z,z_S) &= \frac{F(1-z-n-z_S)(n+z) - n(1-n-z) }{(1-z-z_S)(1-z-n + F(n+z))},\\
        \beta(n;N,F,z,z_S) &= \frac{1}{N} \frac{F(1-z-n-z_S)(n+z) + n(1-n-z)}{(1-z)(1-z-n + F(n+z))}.
    \end{split}
\end{equation}

Note that as $N\rightarrow\infty,$ $z_S\rightarrow0$: in this limit Eqn.\ \ref{alphabeta} is identical to Eqn.\ \ref{alphabeta_new}.

\section{Stability of Fokker-Planck Solutions}

The function $\alpha(n)$ given above represents the difference between the probability of the proportion of weak opinion in the graph increasing and decreasing: the sign of $\alpha(n)$ therefore describes the deterministic direction in which the proportion of weak opinion changes over the system. The equilibrium points occur when $\alpha(n)=0$, which occur for 

\begin{equation}
    \begin{split}
        n_1 &= \frac{F z}{1-F}\\
        n_2 &= 1-z.
    \end{split}
\end{equation}

We can intuit the stability of $n_1$ by considering the direction of deterministic flow for $n= n_1 + \epsilon$, where $\epsilon\to 0$. Substituting this in, we obtain:

\begin{equation}
    \begin{split}
        \alpha(n) &= \frac{1-n-z}{(1-z)\left[1-n-z + F(n + z)\right]} \left[F(n+z) - n\right]\\
        &= \frac{1-n_1 - \epsilon-z}{(1-z)\left[1-n_1 - \epsilon-z + F(n_1 + \epsilon + z)\right]} \left[F(n_1 + \epsilon+z) - n_1 - \epsilon\right]\\
        &= \frac{1-n_1 -z - \epsilon}{(1-z)\left[1-n_1 - \epsilon-z + F(n_1 + \epsilon + z)\right]} \left[F(n_1 +z) - n_1 + \epsilon (F-1)\right].
    \end{split}
\end{equation}

We recognise in the last square-bracket term that $F(n_1 +z) - n_1 =0$ by definition. Furthermore, given $\epsilon \ll z$, we take $z+\epsilon\approx z$:

\begin{equation}
    \begin{split}
        \alpha(n) &= \frac{1-n_1 -z}{(1-z)\left[1-n_1 - z + F(n_1 + z)\right]} \left[\epsilon (F-1)\right].
    \end{split}
\end{equation}

To determine the sign, we shall plug in $n_1 = \frac{F z}{1-F}$:

\begin{equation}
    \begin{split}
        \alpha(n) &= \frac{1-\frac{F z}{1-F} -z}{(1-z)\left[1-\frac{F z}{1-F} - z + F(\frac{F z}{1-F} + z)\right]} \left[\epsilon (F-1)\right]\\
        &= \epsilon \frac{F+z - 1}{(1-z)^2}
    \end{split}
\end{equation}

Thus, $n_1$ is stable if $F+z - 1 < 0 \Rightarrow \frac{z}{1-F}<1$. Note that $\frac{z}{1-F}$ corresponds to the total proportion of nodes which subscribe to the weaker opinion (including zealots) when $n = n_1$, meaning that $n_1$ is stable for the physical case that the total proportion of nodes of the weaker opinion is less than $1$: $n_1$ is therefore unstable in the un-physical case when $\frac{z}{1-F}>1$. Correspondingly, $n_2$ is stable when $\frac{z}{1-F}>1$ and unstable when $\frac{z}{1-F}<1$. The stable long-term proportion of nodes subscribing to the weaker opinion, $P^*(z,F)$, is therefore 

\begin{equation}\label{MeanField}
    P^*(z,F)=\min\left(\frac{z}{1-F},1\right).
\end{equation}

\bibliography{bibliography}